# Deep Reinforcement Learning-Based Control Strategy with Direct Gate Control for Buck Converters

Noboru Katayama

*Abstract*—This paper proposes a deep reinforcement learning (DRL)-based approach for directly controlling the gate signals of switching devices to achieve voltage regulation in a buck converter. Unlike conventional control methods, the proposed method directly generates gate signals using a neural network trained through DRL, with the objective of achieving high control speed and flexibility while maintaining stability. Simulation results demonstrate that the proposed direct gate control (DGC) method achieves a faster transient response and stable output voltage regulation, outperforming traditional PWM-based control schemes. The DGC method also exhibits strong robustness against parameter variations and sensor noise, indicating its suitability for practical power electronics applications. The effectiveness of the proposed approach is validated via simulation.

*Index Terms*—DC-DC converter, optimal control, direct gate control, deep reinforcement learning, artificial intelligence

## I. INTRODUCTION

RECENTLY, various approaches have been proposed to apply deep reinforcement learning (DRL) techniques to the control of power electronic converters. Hajihosseini et al. [1] utilized RL to optimize the parameters of a PI controller in a buck-boost converter, demonstrating superior tracking performance to the reference signal compared to terminal sliding mode control.

Previous studies have investigated the modification of the duty ratio of gate signals in switching devices using DRL. For example, Gheisarnejad et al. [2] proposed a method to enhance the control performance of power electronic converters by augmenting the control signal of a conventional PID controller with a compensation signal obtained through deep deterministic policy gradient. Their approach demonstrated improved response speed and stability, as well as enhanced robustness against dynamic variations in the operating conditions. Cui et al. [3] proposed a model-free DRL-based control method to address voltage instability in DC-DC buck converters caused by fluctuations in constant power loads. In their study, the DRL model outputs discrete values representing the change in the duty ratio.

Several studies have investigated the direct control of the duty ratio in power electronic converters using DRL. Lee et al. [4] employed a model trained via the Soft Actor-Critic (SAC) algorithm to directly regulate the PWM duty ratio of a DC–DC buck converter, demonstrating faster transient response compared to conventional methods such as PI control and model predictive control (MPC). Rajamallaiah et al. [5] also applied DRL to control the duty ratio of a buck converter and evaluated multiple DRL algorithms, concluding that Twin Delayed Deep Deterministic Policy Gradient (TD3) outperformed others in terms of control performance. In addition, DRL-based duty radio control strategies for DC–DC converters have also been explored in [6]-[8].

While most prior work focuses on duty ratio modulation per PWM cycle, DRL has shown promise in more complex control problems beyond power electronics, such as robotics and video game environments. Among these studies, only Qashqai et al. [9] directly learned and controlled the on/off states of individual switching devices, rather than computing a duty ratio for PWM generation. Their DRL-based approach, applied to a three-level neutral point clamped (NPC) converter, exhibited robustness against parameter variations in passive components such as inductors and capacitors, and maintained stable operation even under measurement noise.

Building on these insights, it is anticipated that DRL can be employed to directly control the on/off states of switching devices in converter topologies such as buck converters, without relying on PWM generation. This enables more flexible and faster control by eliminating the constraints imposed by a fixed PWM period. Specifically, a control strategy can be envisioned in which circuit voltages and currents are sampled at a higher rate than the switching frequency, and the appropriate switching actions are determined by the DRL agent at each sampling instant without the conventional duty ratio calculation. Since the control target in Qashqai et al.'s study was a DC/AC conversion circuit, the control interval was 20 µs; however, in buck converters, a much higher control frequency is required. Furthermore, a comparative evaluation with DRL-based control methods that output the duty ratio in previous studies is also required.

In this study, a DRL-based control method—hereafter referred to as direct gate control (DGC)—is proposed to directly generate gate signals for a buck converter, allowing the agent to determine the switching state of the MOSFET at each control step without relying on PWM generation. The proposed DGC method is evaluated through simulation studies, demonstrating its effectiveness in achieving voltage regulation while simplifying the overall control architecture.

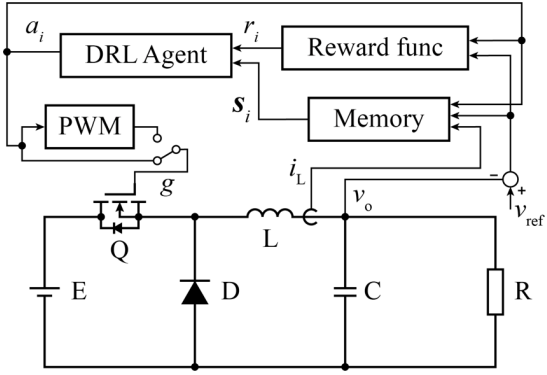

Fig. 1. Block diagram of DRL-based controller for buck converter with direct gate control.

Table I. Circuit parameters of the buck converter.

| Parameter | Value |
| --- | --- |
| Input voltage, $E$ | 20 V |
| Inductance, $L$ | 47 μH |
| Inductor ESR, $R_L$ | 10 mΩ |
| Capacitance, $C$ | 470 μH |
| Capacitor ESR, $R_C$ | 100 mΩ |
| Load resistance, $R$ | 1-∞ Ω |

Table II. Hyperparameters for DRL.

| Parameter | Value |
| --- | --- |
| Discount factor | 0.999 / 0.99 (for PWM) |
| Clip range | 0.2 |
| Learning rate | $3\times 10^{-4}$ |
| Number of epochs | 10 |
| Batch size | 64 |
| Rollout steps | 2048 |
| Train steps | 1,000,000 |
| Policy network | [64, 64] |
| Critic network | [64, 64] |
| Reward parameters, $\alpha, \beta, \zeta, \delta$ | 0.2, 0.004, 0.1, 4 |

## II. METHODOLOGY

*A. System Description*

The target system in this study is a conventional buck converter, consisting of a single MOSFET as the main switching device, a freewheeling diode, an inductor, an output capacitor, and a resistive load as shown in Fig. 1. This topology is widely used in DC–DC conversion due to its simplicity. In this study, the buck converter is selected not only for its simplicity but also to focus on the control aspects of DRL. The primary objective is to regulate the output voltage to a desired reference level by appropriately controlling the switching state of the MOSFET. The nominal circuit parameters, including component values and target output voltage, are summarized in Table I.

To implement and evaluate the proposed control strategy, the converter is modeled and simulated in PLECS, a simulation environment specifically designed for power electronic systems. All components are treated as ideal except for the inclusion of equivalent series resistance (ESR) in the inductor and capacitor. At each control step, PLECS exports the current system state—such as output voltage and inductor current—to the agent for computing the next gate signal.

*B. Control Framework*

The proposed framework eliminates the PWM generation stage and allows the DRL agent to directly determine the gate signal of the switching device in a binary signal. By doing so, the control policy gains more flexibility in responding to system dynamics without being constrained by a predefined modulation scheme. Moreover, this architecture simplifies the overall control structure and opens the possibility for higher-speed and more adaptive switching behavior.

The control period is set to 1 μs, which defines both the state observation interval and the gate signal update rate. At each time step, the current system state—such as output voltage and inductor current—is sampled and passed to the DRL agent, which returns a discrete action representing the next gate state.

*C. Deep Reinforcement Learning Algorithm*

The control policy for the proposed strategy is trained using the Proximal Policy Optimization (PPO) algorithm [10], a model-free, on-policy reinforcement learning method. PPO has gained popularity due to its ability to achieve stable and sample-efficient learning while being relatively easy to implement and tune. It adopts an actor–critic architecture, in which the policy and the value function are represented by separate neural networks and optimized simultaneously. Another reason for selecting the PPO algorithm is its ability to handle both continuous and discrete action spaces. This enables a fair comparison between the proposed DGC method and the conventional PWM-based approach using the exact same algorithmic framework.

In this study, the action space is defined as a binary set representing the on/off states of the switching device. During training, actions are sampled stochastically from the policy distribution. During evaluation, a deterministic policy is used by selecting the action with the highest probability (i.e., the most probable action). This ensures reproducible behavior when comparing performance with baseline methods.

The neural networks for both the policy and value functions consist of two fully connected layers, each followed by a ReLU activation function. The output layer of the policy network produces the logits for a categorical distribution over the two discrete actions. The hyperparameters used for PPO training are summarized in Table II.

*D. Definition of State, Action, and Reward*

The control problem is formulated as a discrete-time Markov decision process, in which the agent observes the current system state, selects a control action, and receives a scalar reward at each control interval.

The state vector at each time step includes the recent history of the voltage error between the converter output and the reference voltage, the instantaneous inductor current, and the switching device's on/off states, each recorded on the memory over the previous ten sampling steps. This temporal history is incorporated to provide the agent with short-term dynamic information relevant to recent switching behavior and state transitions, thereby enhancing the observability of internal converter dynamics.

The action space is defined as a discrete set with two possible values, on and off states of the switching device. At every control step, the agent selects one of the two actions. To account for the processing delay introduced by the ADC and the policy network inference, the selected action is applied to the switching device with a one-step delay.

The reward function is defined to promote accurate output voltage regulation and penalize excessive switching. A general form of the reward used during training is given by:

$$r_1 = \frac{\alpha}{|v_{err,t}| + \epsilon} - \zeta|v_{err,t}| - \beta \quad (1)$$
$$r_2 = -\delta|a_t - a_{t-1}| \quad (2)$$
$$r = r_1 + r_2 \quad (3)$$

where $v_{err,t}$ and $a_t$, denotes the output voltage error and switching state at time step $t$, respectively. The other constants are listed in Table II. The behavior of circuit controlled by DRL can be flexibly designed through the reward function.

*E. Training Environment*

Each training episode consists of 2,000 time steps, corresponding to a simulation duration of 2 ms with a control interval of 1 μs. To promote generalization and ensure the agent experiences a wide range of operating conditions, the initial state for each episode is randomized. Specifically, the initial values of the inductor current, output capacitor voltage, and load resistance are independently sampled from predefined ranges. This setup allows the agent to encounter diverse transient and steady-state behaviors during training.

*F. Baseline for Comparison*

To evaluate the effectiveness of the DGC control, a baseline method is implemented in which the PWM duty ratio is directly controlled by a DRL agent. The same PPO algorithm and neural network architecture and the state vector are used for the baseline to ensure a fair comparison. In the baseline configuration, the agent outputs a continuous-valued duty ratio at each control step. This value is then applied to a conventional fixed-frequency PWM generator to produce the corresponding gate signal. The PWM period and control interval are set to 10 μs. Since the control interval is longer than the DGC method by ten times, the reward value is multiplied by 10 to account for the difference in control frequency. The discount rate is also adjusted to 0.99 to maintain consistency with the DGC method.

## II. RESULTS AND DISCUSSION

*A. Dynamic Response to Load Step Changes*

The fundamental behavior of the proposed DRL-based gate signal control (DGC) was first verified under nominal conditions. Fig. 2 compares the response of the proposed method (blue line) and the conventional PWM control (orange line) under a load step disturbance from 15 Ω to 1 Ω at 0.0 ms. The top two graphs depict the gate signal and duty ratio, the middle plots show the inductor current, and the bottom plots display the output voltage (solid line) with the reference voltage (dotted red line). Since the concept of a duty ratio does not directly apply to DGC, an effective duty ratio, calculated by applying a digital low-pass filter to the gate signal, is shown for illustrative purposes.

After the output voltage initially drops to approximately 13.8 V due to rapid load current increase, the DGC method quickly adapts by adjusting the gate signal. The output voltage starts to rise at 0.06 ms, reaching the reference voltage within 0.15 ms. The gate signal stays high for 0.14 ms after the load change, subsequently, the gate signal alternates between on and off, bringing the output voltage asymptotically toward the reference voltage. Notably, the gate signal transitions are not constrained by a fixed switching frequency, providing greater control flexibility.

However, the PWM control method exhibits a slower response. The output voltage drops to 13.2 V and recovers to the reference voltage after 0.2 ms. The output voltage overshoots to 15.2 V and oscillates before settling around 15 V. The duty ratio is fixed at 1.0 for 0.13 ms, and adjusted to settle the output voltage to the reference voltage. Although the response is adequate, the performance shows poorer compared with DGC due to control delay and the fixed switching frequency.

*B. Robustness Evaluation Under Parameter Variations*

To evaluate the robustness of DGC against parameter variations due to its flexible control scheme, dynamic response was evaluated varying important circuit parameters such as inductance and equivalent series resistance (ESR) of the output capacitor. The inductor value is typically affected by the manufacturing process, while the ESR of the output capacitor can change due to aging or temperature effects.

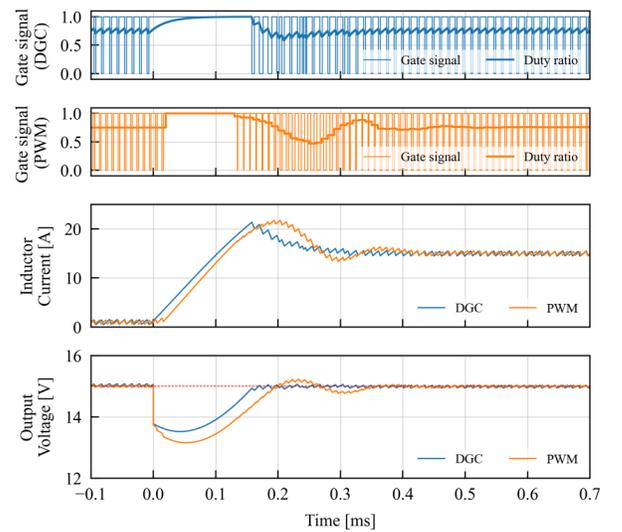

Fig. 2. Comparison between the proposed DGC method and conventional PWM control in response to a load step change under nominal circuit parameters.

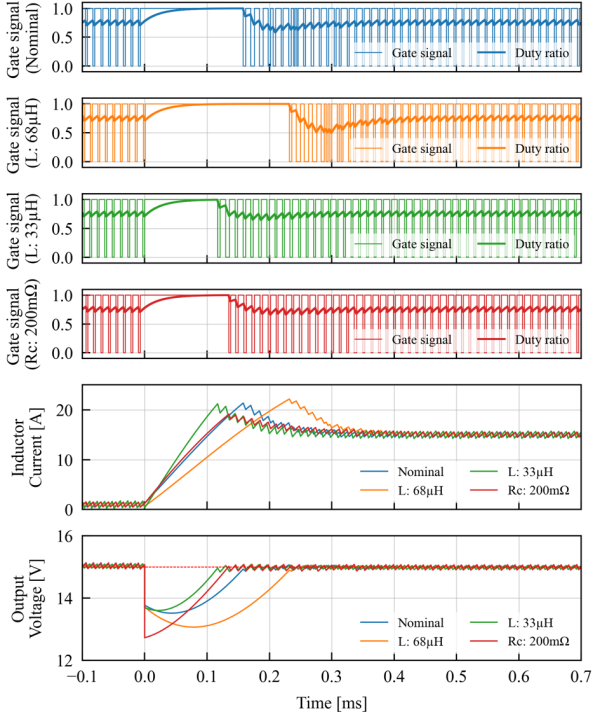

Fig. 3. Response under variations in inductance and ESR of the capacitor.

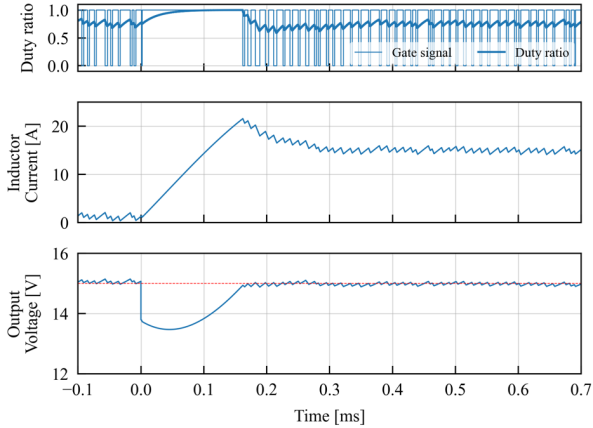

Fig. 4. Response under nominal circuit parameters with injected measurement noise.

Fig. 3 shows the output voltage response when the inductor value is changed to 33 μH and 68 μH, and the ESR of the output capacitor is doubled from 100 mΩ to 200 mΩ, respectively. For all cases, the output voltage recovered to the reference voltage in a stable manner. Although the recovery time is slightly changed for replaced circuit parameters, all the cases show that the output voltage remains well-regulated.

*C. Robustness Against Sensor Noise*

Sensor noise was introduced to the output voltage measurement to evaluate further robustness. The noise was modeled as a Gaussian white noise with a standard deviation of 0.01 V for the output voltage and 0.1 A for the inductor current. Fig. 4 shows the output voltage response under this noisy condition. The DGC method successfully maintained stable output voltage regulation despite the presence of noise, demonstrating robustness against measurement uncertainties. The switching behavior responds to the noisy measurements, showing the agent's ability to adapt quickly to perceived changes in the output voltage while maintaining overall stability.

## II. CONCLUSION

This study proposed a novel direct gate control (DGC) method for buck converters using deep reinforcement learning. The DGC method directly generates gate signals for the switching device, eliminating PWM generation. Simulation results demonstrated that the DGC method achieves fast transient response and stable output voltage regulation, outperforming traditional PWM-based control methods. Additionally, the DGC method exhibited robustness against parameter variations and sensor noise, confirming its potential for practical applications in power electronics.